# Digital payment policy impact analysis on the intention to use QRIS (quick response code Indonesian standard) during COVID-19 pandemic[1]

WISHNU BADRAWANI[a2]

[a]Bank Indonesia, Indonesia

**Abstract:**

This study aims to evaluate the adoption of Bank Indonesia's QRIS (Quick Response code Indonesian Standard) payment system policy. The evaluation is hindered by the contemporaneous emergence of the COVID-19 pandemic, which acts as a confounding factor in adopting the new payment instrument. To disentangle the impact of central bank policy from the pandemic, a novel variation of the model of Unified Theory of Acceptance and Use of Technology (UTAUT) is proposed and is estimated using purposive sampling from an online survey with 572 respondents during the pandemic. The result of the study successfully disentangles the policy effect from the pandemic effect, and also separate the risk of pandemic with common risks (PR) and other technology adoption determinants. The results indicate that perceived central bank policy and pandemic risk are the most influential variables affecting the intention to use QRIS. The findings suggest that this measurement approach can be appropriately used as a complementary tool to examine the effectiveness of the central bank's policy in influencing people's behavior.

---

[1] Presented at International Conference On Economics, Business And Economic Education (ICE-BEES) 2021, Semarang City Central Java Province, Indonesia
[2] Corresponding author: w_badrawani@bi.go.id

# INTRODUCTION

Measuring policy impact is critical for an organisation; it helps increase stakeholder and public awareness of its relevance (OECD, 2015). Furthermore, policy impact evaluation helps an organisation determine whether it is accomplishing its long-term objectives and goals, as well as identify and utilise its most valuable resources. One way of studying the impact of regulatory policy is by examining behavioural changes in the audience, see, for example, Coglianese (2012). In addition to its principal role as a monetary authority, Bank Indonesia[3] is mandated by the Indonesian constitution to implement macroprudential and payment system policies. The impact of a macroprudential policy has been studied extensively see, e.g. Tressel and Zhang (2016). However, research on payment system policies is surprisingly limited, both in terms of understanding their transmission channels and evaluating their effectiveness, as Gogoski (2012) mentioned.

In late 2019, Bank Indonesia launched a national standard for quick response (QRIS.id) codes in the payment system called the QRIS. QRIS refers to an advanced QR pattern for payments that allows contactless transactions and interoperability among electronic money and mobile banking providers. Previously, consumers had to subscribe to many payment service providers or platforms to settle their online transactions, while merchants had to apply many QR codes from different service providers, which further required additional hardware in the form of electronic data capture (EDC) machine or an electronic funds transfer at the point of sale (EFTPOS) terminal. The aim of QRIS was to connect the multitude of individual platforms with a single and efficient system through which all online transactions would take place.

This paper aims to evaluate the impact of central bank policy on the behavioural intention of people in adopting QRIS that subsequently increase the number of non-cash transactions. However, policy impact evaluation is a challenging task as, concurrently with the introduction of QRIS, the COVID-19 pandemic emerged and brought containments and economic activity restrictions and therefore, it is difficult to separate the effect of central bank policy from the impact of the pandemic. People were encouraged to spend more time at home during the pandemic due to government-enforced lockdowns, avoiding shopping and dining in crowded spaces. Cashless payments were promoted to limit the risk of infection (WHO, 2020), despite the fact that these payment instruments are still not widely available in many countries (Auer et al., 2020). As a result, the usage of fintech products and services, such as digital payments and remittances, increased significantly during the COVID-19 pandemic (Rowan et al., 2021) and is often higher in countries with more stringent COVID-19 containment measures.

The effect of the Covid-19 pandemic on people's intention to adopt new technology, such as the digital wallet and mobile payment, has been studied, see, e.g. Aji et al. (2020) and Zhao and Bacao (2021). Also, previous research has emphasised the impact of government policies on people's intentions to use new technologies. For example, Chong et al. (2010) discovered that government endorsement such as a clear regulation substantially alters the intention to use internet banking; and government support significantly impacts farmers to adopt new technology initiated by the government (Mandari et al., 2017). However, to the best of our knowledge, no previous study has evaluated and distinguished the influence of government (central bank) policy and the Covid-19 epidemic on people's behavioural intention in using a new payment instrument, particularly in the Asian country region.

---

[3] The central bank of Indonesia

Accordingly, this research aims to conduct an empirical examination of the effectiveness of Bank Indonesia's policy in influencing people's behaviour intention toward the adoption of QRIS in the context of the COVID-19 pandemic. Policy effectiveness refers to the degree to which QRIS impacted the targeted behaviour, as in Coglianese (2012), and is measured using a novel model of the Unified Theory of Acceptance and Use of Technology (UTAUT) originally proposed by Venkatesh et al. (2003). The novelty in this paper is that additional factors are introduced, namely perceived risk (PR), force majeure (FM) and perceived government role or law enforcement (LE), which allow the model to disentangle central bank policy effectiveness from the impact of the pandemic. The model development and the specific interest of the study will be considered the novelty of this research.

The research is based on a purposive sample of 572 observations of consumers and vendors in many cities in Indonesia. Based on the data set, the extended UTAUT model is estimated using the methods of a partial least squares (PLS) method to structural equation modelling (SEM). The findings show and disentangle the role of the central bank's policy and the influence of the COVID-19 pandemic on people's behaviour intention to adopt QRIS. As a result, this study provides an alternative method for policy evaluation for central banks or government institutions, as well as input for players in the payment system industry to focus on the elements that will increase the adoption of new payment instruments.

The remainder of the paper is structured as shown below. The second section reviews the available literature and developed hypotheses, while the third section describes the methodology. The empirical results and discussion are presented in sections four and five, followed by the conclusion in section six.

## LITERATURE REVIEW AND HYPOTHESES DEVELOPMENT

### QRIS (Quick Response Code Indonesian Standard)

As part of the Indonesian payment system blueprint, Bank Indonesia and the payment system association developed a standardised QR code for all payment service providers in Indonesia called QRIS. It aims to promote interoperability and enhance efficiency by allowing consumers to transfer funds to their counterparts who use different payment services (Bank Indonesia, 2019). In addition, QRIS enables contactless payment for server-based e-money or digital wallets and mobile banking. Accordingly, QRIS refers to a QR code payment standard for Indonesian payment instruments that allows a user of one payment service to transfer funds to their counterpart who uses a different payment service provider. The central bank also applied a rigorous licensing process of QRIS members with a recommendation from the standard agency as a requirement and surveillance and sanction imposition.

Despite the ineffectiveness of QRIS implementation among merchants in Semarang city (Putri, 2020), 6 million subscribers were registered in 2020 throughout 34 provinces and 480 districts, connecting users of 50 brands of digital wallets and more than 20 mobile banking brands throughout the country (Bank Indonesia, 2021). According to Donovan (2012), mobile banking or digital wallets have a number of advantages over card payment (debit and credit cards or electronic money) and cash, including being significantly more cost-effective, safer, and convenient than cash or card-based e-money, as well as increasing financial access for the poor.

Regardless of its merits, Camner (2013) argued that a segregated and isolated payment environment is unlikely to sustain significant usage and may even result in a monopolistic market. This finding is corroborated by Banda et al. (2015), who reveals that high concentration levels in a specific industry may lessen the incentive to innovate and lead to high prices. As a result, the authority needs to promote fair competition that fosters innovation and consider the likelihood of anti-competitive behaviour that results in economic inefficiency (Macmillan et al., 2016), which supports the QRIS initiation. Additionally, this form of initiation is typically accounted for as part of a central bank's policy; for example, see, Khiaonarong (2003).

**COVID-19 and Transaction Behaviour**

COVID-19 has profoundly impacted lives, livelihoods and the global economy since it was declared as a global pandemic by the World Health Organization in March 2020. According to the World Health Organization (2020), COVID-19 can be transmitted directly or indirectly through contact with sick individuals or contaminated goods or surfaces. As a result, many governments have responded by introducing social distancing regulations, temporary workplace and school closures, and social lockdowns to combat COVID19's harmful health consequences.

These containment policies had severely impacted society and altered consumer's behaviour. As a result, people were forced to spend more time at home, could not directly contact coworkers, and do more online. Consequently, demand for digital financial services and mobile money surged (Agur et al., 2020) while demand for conventional currency decreased (Cevik, 2020). In this context, shifting consumer behaviour in response to COVID-19 has boosted the growth of contactless instruments like mobile payment and digital wallets while limiting COVID-19 transmission (Pal and Bhadada, 2020).

In Indonesia, the government has encouraged businesses, food vendors, transportation providers, and traditional marketplaces to use non-cash transactions (Bank Indonesia, 2021). Additionally, the central bank has initiated numerous programmes to accelerate the transition to a cashless economy and to foster interoperability of contactless payment instruments (i.e. e-wallet and mobile payment) between the fintech industry and banking by establishing the QRIS, as recommended by the Bank for International Settlements (2020).

The emergence of the pandemic has made it difficult for researchers to evaluate the impact of policies that were introduced contemporaneously. Many traditional methods based on secondary data are difficult to apply, for example, difference-in-differences (Lechner, 2011, Fredriksson and Oliveira, 2019). Therefore to avoid this difficulty, we employ a novel version of the UTAUT model of Venkatesh et al. (2003), which is capable of separating policy from the pandemic.

**Research framework**

This study examines how central bank policy influences people's intention to use QRIS during the COVID-19 outbreak using the UTAUT model. Venkatesh et al. (2003) established the Unified Theory of Acceptability and Use of Technology (UTAUT), which provides measurement methods for assessing user acceptance of new technology products. UTAUT emerged from earlier theories defining user acceptance of information technologies and has been empirically validated, albeit the outcome may vary by field of research and country (Attuquayefio and Addo, 2014). According to the UTAUT model, four latent factors influence behaviour intention (BI) in adopting new technology: performance expectation (PE), effort expectation (EE), social influence (SI), and facilitating condition (FC). For that reason, UTAUT is the appropriate theory for studying QRIS acceptability since it enables the examination of variables affecting the information technology (IT) surrounding the QRIS while also taking social issues into account. Additionally, the UTAUT model has been empirically validated in numerous research relating to information technology in many countries, see, e.g. Williams et al. (2015).

We extend the original UTAUT model in this study to account for the impact of the COVID-19 pandemic and the central bank policy on the behavioural intention to adopt new technology, specifically QRIS. We also include variable perceived risk (PR) to distinguish risks associated with the COVID-19 pandemic from common risks related to accepting new technology. The conventional UTAUT constructs, namely performance expectancy, effort expectancy, social influence, and facilitating condition, were combined with new latent variables, including perceived risks (PR), the COVID-19 pandemic or other unprecedented events referred to as force majeure (FM), and perceived government (central bank) policy referred to as law enforcement (LE). As a result, as illustrated in Figure 1, the proposed research model is developed.

**Performance expectation (PE)**

The term *Performance Expectation* refers to the extent to which users anticipate that adopting new technology would enable them to perform their jobs more effectively (Venkatesh et al., 2003). PE has been shown empirically to be a significant predictor of behavioural intention to adopt new mobile payment technologies (Morosan and DeFranco, 2016), particularly for organisational users (Venkatesh et al., 2012).

Additionally, its significance in influencing intention to adopt new technology such as a digital wallet or mobile payment also appeared during the COVID-19 pandemic, see, e.g. Aji et al. (2020) and Zhao and Bacao (2021). As a result, a rise in PE will have an effect on the user's behaviour intention and, ultimately, on the new technology's acceptability. Therefore, the following hypotheses were formed:

Hypothesis 1 (H1): Performance expectancy positively affects behavioural intention (BI) in using QRIS.

*Effort Expectancy (EE)*

The term *Effort Expectancy* refers to the easiness level associated with individuals operating the new technology (Venkatesh et al., 2003). EE approximates the perceived ease of use in the technology acceptance model (TAM) proposed by Davis (1989) or the adverse value of complexity in innovation diffusion theory (IDT) proposed by Rogers (1983), which both explain people's belief that using a given technology will be effortless. This variable has been consistently publicised as a critical predictor in explaining intention's behaviour, see, e.g. Venkatesh et al. (2016). According to Aydin and Burnaz (2016), ease of use appears to be the most significant predictor of consumers' mobile wallet usage. As a result, we hypothesised as follows:

Hypothesis 2 (H2): Effort Expectancy (EE) positively affects behavioural intention (BI) in using QRIS.

*Social Influence (SI)*

Venkatesh et al. (2003) defined *Social Influence* as the extent to which users believe that people surrounding them think they should use new technology. According to the theory of reasoned action (TRA) (Fishbein et al., 1980) and the theory of planned behaviour (TPB) developed by Ajzen (1985), it is represented as a Subjective Norm, and it has been empirically demonstrated to have a direct impact on behavioural intention (Venkatesh et al., 2012). Social influence or subjective norm factor also found to have a direct effect in altering people's behaviour intention to use near field communication mobile payments (NFC-MP) and internet banking, see, for example, Morosan and DeFranco (2016) Morosan and DeFranco (2016) and Lee (2009). According to Lin et al. (2019), social influence has a greater impact on the behavioural intention to use mobile payment in Korea than in China due to the differences in demographic characteristics and payment patterns between the two countries; whereas Sun et al. (2012) confirm the effect of subjective norms on the intention to adopt new technology in relation to religious affiliation. Therefore, the hypothesis is proposed as follows:

Hypothesis 3 (H3): Social Influence (SI) positively affects behavioural intention (BI) in using QRIS

*Facilitating Condition (FC)*

*Facilitating Condition* refers to the stage at which a person has confidence in the resources and technical assistance available to support them using the new technology (Venkatesh et al., 2003). Using mobile banking or a digital wallet properly, for example, requires a number of apparatus and supporting ecosystems, such as a server, legal aspects or licences, a network of agents, easy access for user assistance, payment protocol, and security, among others (Gupta, 2013, Sikri et al., 2019). Numerous further studies have established that the facilitating condition plays a significant role in the behavioural intention to use new technology, see, for example, Morosan and DeFranco (2016) in the case for NFC mobile payment in hotels and Widodo et al. (2019) for the adoption of the digital wallet, among other studies. The greater the number of facilities made available by the service provider, the higher the likelihood of adopting new technology. As a result, the following hypothesis is proposed:

Hypothesis 4 (H4): Facilitating Condition (FC) has a positive relationship with the behavioural intention (BI) in using QRIS.

**Perceived risks (PR)**

*Perceived Risk* refers to an individual's perceptions of potential losses associated with pursuing the desired outcome through the use of technology, following Featherman and Pavlou (2003). They classified perceived risk into seven dimensions: performance, financial, time, psychological, and social, as well as privacy and overall risk. For the purposes of this study, we measure perceived risks in accordance with Lee (2009), who pointed out that the amount of intention to use mobile banking is mostly driven by security or privacy risk as well as financial risk. Financial risk, or the possibility of losing money, is a critical matter to consider in the payment sector. Concerns about the possibility of losing money when conducting transactions using mobile banking remains a major concern in developing countries, particularly in Africa and Asia, see, for example, Achieng and Ingari (2015) and Bansal and Bagadia (2018). Concerns about *privacy security* have been raised as a result of the development of mobile money and its rapid adoption (Harris et al., 2012), which should be of concern to issuers and regulators. The elevation of negative perception concerning privacy threats emerges with the increase of online users' competence (Hoffman et al., 1999). In general, perceived risk has an adverse influence on the intention to use mobile banking (Bansal and Bagadia, 2018), mobile payment (Yang et al., 2015), and digital wallet (Liébana-Cabanillas et al., 2020), among other financial services. As a result, we proposed the following hypotheses:

Hypothesis 5 (H5): Perceived risk negatively influences behavioural intentions to use QRIS.

**Force Majeure (FM)**

In this study, the role of the COVID-19 pandemic is captured by adding a new external factor representing the pandemic, namely, *Force Majeure* (FM). There have been various studies on technology acceptance conducted during the COVID-19 pandemic in numerous fields, such as Riza (2021) in Islamic mobile banking and Sukendro et al. (2020) in e-learning platforms among students of sports science education, both of which have been published recently. However, it is arduous to find discourses that explain the impact of the COVID-19 pandemic on the behavioural intention to use technology. In the study of e-wallet usage conducted by Aji et al. (2020), the COVID-19 pandemic is represented as a perceived risk factor that positively affects the intention to use an e-wallet, both directly and indirectly. However, this finding contradicts other studies that indicate that perceived risk had no effect on people's intention to adopt FinTech applications during the COVID-19 epidemic, see, e.g. Nawayseh (2020).

Taking into account the preceding findings, this study attempts to differentiate between common risk factors associated with the acceptance of new technology, as represented by the perceived risk (PR), and risks associated with the COVID-19 pandemic or a situation in which the research was conducted in the manner used in previous studies, by treating the COVID-19 pandemic as an exogenous latent variable, namely force majeure (FM). Thus, force majeure is defined as the user's belief about the possibility of avoiding the negative impact of unprecedented events, namely the COVID-19 pandemic (or disaster), through the application of new technology. Hence, the hypothesis is proposed as follows:

*Hypothesis 6 (6):* There is a positive relationship between Force Majeure (FM) and behavioural intention (BI) in using QRIS.

**Law Enforcement or Perceived Government's Policy (LE)**

A study conducted by Chong et al. (2010) found that government support significantly influences the intention to use internet banking. In the case of Fintech services, this finding is verified by Chen et al. (2019), who stated that government endorsement of legitimacy and reliability would help to increase the public awareness of their new technology. Carter and Bélanger (2005) found that, in the case of new technology introduced by the government, such as e-government, the perceived trustworthiness of the government has a considerable impact on the user's intention to adopt the state's e-government services. Janssen et al. (2018) considered it to be the ultimate predictor, particularly faith in the government, rather than the technology under observation, namely e-government. This finding is consistent with Teo et al. (2008), who argue that individuals' impressions of the quality of an e-government website are influenced by their trust in the government, not its technology.

In the study conducted by Mandari et al. (2017) regarding farmers' behavioural intention towards using m-government services, government support was discovered as a significant factor influencing the intention to use such new technology. Farmers are more likely to accept m-government if they believe the government would deliver benefits related to the latest technology being used. The findings were corroborated by Aji et al. (2020), who explicates that government support indirectly influences the intention to use e-wallets and that the perceived usefulness factor fully mediates this relationship. Thus, without the sense of benefits, government assistance will have no effect on the intention to utilise such technologies. In this study, law enforcement is described as the public's belief and trust in the government's or central bank's new policy, in this case, the introduction of QRIS with any regulations attached to it. As a result, we propose the following hypothesis:

Hypothesis 7 (H7): There is a positive relationship between perceived government's (central bank) policy or law enforcement (LE) and behavioural intention (BI) in using QRIS.

**Insert Figure 1 here**

## RESEARCH METHOD

**Data collection and sampling technique**

This is a cross-sectional study conducted in Indonesia from February to March 2021 using an online Qualtrics survey with self-administered questions. The online survey method was chosen to minimise measurement error caused by the interviewers and to avoid the risk of coronavirus infection. Regarding the ethical consideration of the study, the research was approved by the University of Birmingham's Humanities and Social Sciences Ethical Review Committee.

Prior to the main data collection, a pilot test was conducted to ascertain the clarity of the questionnaire, as suggested by Pickard (2013). It is to ensure that the questionnaire was accurately translated from its original source and did not cause misinterpretation, all questions were easily answered by respondents, and the results were comfortably recorded. A non-probability sampling method applying the purposive sampling technique with judgmental sampling applied in this study ensures only relevant respondents participate in the research (Saunders et al., 2009) who have experience with QRIS.

**Items measurement**

The questionnaire was structured in three sections. The first section informed the respondent about the topic and purpose of the research, followed by a request for electronic consent. The second segment begins with a screening question to determine respondents' relevance, followed by closed-ended questions concerning respondents' sociodemographic factors, such as gender, age, education, and QRIS experience. The third and final section contains close questions that represent indicators of latent variables developed by Venkatesh et al. (2003, 2012) and the aforementioned extended factors to address the stated research objectives. It consists of 23 measurement items that serve as an indicator for eight latent variables: performance expectancy (PE), effort expectancy (EE), social influence (SI), facilitating condition (FC), perceived risk (PR), perceived COVID-19 pandemic or force majeure (FM), perceived government (central bank) policy or law enforcement (LE), and behavioural intention (BI).

All questions use a five-point Likert scale (from 1 to 5, representing "strongly disagree" to "strongly agree"). Appendix A shows a description of variable operationalisation, while Figure 1 illustrates the study model and proposed hypothesis previously addressed.

**Partial Least Squares (PLS)**

For a variety of reasons, a partial least squares (PLS) method to structural equation modelling (SEM) was used with SmartPLS 3.3 (Ringle et al., 2015) to conduct data analysis in this study. To begin, the primary objective of this research was to explore the determinant factors that influence the intention towards the adoption of QRIS in the context of the Covid-19 pandemic and innovation initiated by the government, developed from previous theory, namely UTAUT. Hair Jr et al. (2016) pointed out that the variance-based PLS approach to structural equation model (SEM) offers an alternative to covariance-based SEM that is mainly used for exploratory research and theory development, and it is also suggested for a confirmatory study, see, e.g. Afthanorhan (2013). Second, the causal-predictive aspect of PLS-SEM imposes minimal requirements on distribution normality and sample size Ali et al. (2018).

Hair Jr et al. (2017) also affirms that PLS prediction-oriented approach to SEM provides a prominent predictive accuracy that validates the proposed model and verifies a well-developed causal relationship, even when the data is not normally distributed. In short, PLS-SEM estimates coefficients that maximise the R Square values and minimise the error terms of the endogenous constructs (Hair Jr et al., 2016).

## RESULTS

**Respondent's Demographic Profile**

The survey obtained 849 responses via a web-based online Qualtrics survey; nevertheless, some respondents did not complete the survey (78), eight respondents did not consent, and a significant number of respondents failed the screening question (277 participants). As a result, this study can analyse only 572 valid data points representing 31 Indonesian provinces. It consists of 239 males (41.78%) and 333 females (58.22%), with an average age of 26 to 35 years and a majority of participants holding a bachelor's degree (364 or 63.64 per cent).

The respondents predominantly live in the capital city of a province (207 or 36.19%), followed by district area, and Jakarta and surrounding (Jabodetabek), which accounted for 34.44% and 29.37%, respectively. Most respondents reported having internet access in their premises (475 or 83.04%), while only 97 respondents (16,96%) do not have internet access. Additionally, the majority of respondents indicated that they had used QRIS for more than one year (50.70%). Table 1 shows the demographics of the participants in this study.

**Insert Table 1 here**

**Measurement Model Assessment**

The first stage in evaluating PLS-SEM results is testing the measurement models that consist of reliability (or internal consistency), convergence and discriminant validity (Hair Jr et al., 2016). The reliability test addresses the consistency level of research measurements. Cronbach's alpha and composite reliability were used to assess the construct's reliability; the higher values mainly represent higher levels of reliability with a minimum threshold value of 0.7 (Kline, 2015). As shown in Table 2, all Cronbach's alpha and composite reliability values of latent variables exceed the 0.70 suggested, indicating that the reliability of the constructs has been presented.

Convergent validity indicates the extent to which the assessments under each construct measure the same attribute. The validity test result of all indicator variables in this questionnaire shows that loading factors are higher than 0.40, except for the observed variable FC4, meaning this indicator variable needs to be eliminated. All items have a loading above the suggested threshold of 0.708 (Hair Jr et al., 2016). Then, we examine the convergent validity in the construct level using the average variance extracted (AVE). AVE in each construct was above the 0.5 thresholds, as Hair et al. (2019) suggested. The result of the validity test is shown in Table 2.

**Insert Table 2 here**

The discriminant validity test represents the extent to which a proposed construct is empirically dissimilar from other constructs in the given model. It is measured using the Fornell-Larcker criterion that compares the square root of the AVE of each construct with the correlations of the latent variables (Hair Jr et al., 2016) and the Heterotrait–Monotrait (HTMT) ratio of the correlations (Henseler et al., 2015). The result of discriminant validity shows that the square roots of all AVE were much more significant than correlations among constructs (Table 3), and the HTMT result successfully meets the threshold of maximum 0.9, thereby satisfying discriminant validity. Additionally, the proposed model's Standardised Root Mean Squared Residual (SRMR) was reported to be 0.045 (less than 0.08), further evidence to the appropriateness of the composite factor model proposed fits the collected data (Henseler et al., 2014, Hair Jr et al., 2016). The result of the discriminant validity test is available in Table 3 and Table 4.

**Insert Table 3 here**

**Insert Table 4 here**

**Structural Model Assessment**

PLS-SEM is different from CB-SEM; it estimates the parameters to maximise the explained variance of the endogenous constructs. Hence, according to Hair Jr et al. (2016), examining the structural model is primarily based on heuristic criteria influenced by the model's predictive capabilities rather than testing goodness-of-fit.

In this study, we employed a bootstrapping procedure with 5,000 resamples for the significance testing, and the sign of the parameters was consistent. The statistical test results show that the value of $R^2_{adj.}$, coefficient of determination adjusted, is 0.640. It indicates that the variance of exogenous latent variables can explain the variance of the Behavioural Intention in using QRIS by 64.0%. The value of $Q^2$, predictive relevance, is 0.561, which suggest that the exogenous constructs have considerable predictive relevance in explaining the appointed endogenous variable, namely Behavioural Intention. The effect sizes ($f^2$), the relative impact of an omitted exogenous variable from the model on the endogenous variable, were reported as a part of the structural model assessment. Hair et al. (2019) pointed out that the effect size examination should follow Cohen (2013) guidelines; the values of 0.02 refer to small, and the value of 0.15 and 0.35 refer to moderate and high $f^2$ effects sizes, respectively. This guiding principle also applies in examining the relative impact of predictive relevance or the $q^2$ effect size.

**Insert Table 5 here**

The results of the structural model in Table 1 show that all hypotheses are supported. The hypothesis testing result indicates that performance expectation ($\beta = 0.096$; $p < 0.05$; $f^2 = 0.012$, $q^2 = 0.008$) and effort expectation ($\beta = 0.096$; $p < 0.05$; $f^2 = 0.009$, $q^2 = 0.008$) were shown to have a significant positive influence on behavioural intention to use QRIS, however with no significant effect size; thereby, supporting H1 and H2. Social influence was found to significantly affecting the intention to use QRIS with a positive sign with small effect size ($\beta = 0.185$; $p < 0.01$; $f^2 = 0.049$, $q^2 = 0.035$) followed by facilitating condition ($\beta = .173$; $p < 0.01$; $f^2 = 0.037$, $q^2 = 0.026$); hence, supporting H3 and H4.

The behavioural intention of adopting QRIS during the COVID-19 pandemic is most significantly influenced by perceived central bank's policy or law enforcement (LE) with a positive sign (ß = 0.227; p < 0.001; $f^2$ = 0.074, $q^2$ = 0.053), followed by perceived COVID-19 pandemic or force majeure (FM) (ß = 0.203; p < 0.01; $f^2$ = 0.057, $q^2$ = 0.040). Thus, hypotheses H6 and H7 are validated, respectively. Moreover, perceived risk was found to gain a significant negative influence on intention to use QRIS (β = -0.060; p < 0.01; $f^2$ = 0.009, $q^2$ = 0.005), although it has no effect size, which supports H5. Accordingly, it shows that the predictors of the UTAUT (Venkatesh et al., 2003) used in our proposed model are still reliable to predict the behavioural intention toward new technology adoption, particularly for the use of QRIS in Indonesia.

## DISCUSSION

The results of our study indicate that the COVID-19 pandemic (FM) significantly affected customers' intentions to use QRIS. These findings corroborate those of Strielkowski (2020), Horgan et al. (2020), and Sutarsa et al. (2020), who pointed out that COVID-19 has significantly helped accelerate the transformation of people behaviour in adopting new technology, such as the use of e-health platform, digital revolution in academia and higher education, and digital payment instruments usage. Notably, the contactless characteristic of the QRIS payment is favourable for a user in retaining social distancing and protecting personal safety during Covid 19 pandemic.

The outcomes of this study reveal that the perceived central bank's policy (LE) of introducing a new payment platform called QRIS in the country has greatly influenced people's intent to use it, followed by the COVID-19 pandemic (FM). The result corroborates Janssen et al. (2018) and Carter and Bélanger (2005), who explicates that the perceived trustworthiness and credibility of the government is a crucial factor that determines the behaviour and intention to use the new services promoted by the government, as well as the expectation of benefit related to the new technology proposed Aji et al. (2020). While being a significant determinant, the perceived risk (PR) component does not become a concern of participants when adopting QRIS in this study (it has no effect size). These findings may be related to the transparent and rigorous process of licencing for QRIS member approval with the dual-stage licensing process involving both the central bank and standard agency. This process not only reduces the potential risks but subsequently helps establish and maintain the credibility of the central bank policy.

Performance expectation (PE) was found to influence the intention to use QRIS significantly; however, it has no effect size. This outcome may be attributed to the unique characteristic of QRIS as a national QR code standard that facilitates the interoperability of payment services that were previously fragmented. QRIS provides a new platform that mediates and settles transactions between consumers and merchants with different payment service providers in real-time verification, with no funds to be held. Additionally, the number of recorded complaints is negligible until this study is complete (Mediakonsumen, 2020).

This finding holds for the effort expectation (EE) as well. Although EE was determined to be a statistically significant variable, it has no effect size. This result is consistent with the findings of Riza (2021) and Kadim and Sunardi (2023); however, there is a discrepancy in the observed effect sizes. The observed discrepancy in the effect size may be attributed to the inclusion of additional variables in the UTAUT model to account for and distinguish the impact of COVID-19 and the central bank's policy. This factor was not considered in the analysis conducted by Kadim and Sunardi (2023), although QRIS was the same object being observed. QRIS has a unique feature as a payment channel; it does not need an application to be installed or an initial cost for infrastructure installation, such as an EDC machine or EFTPOS terminal. It is automatically added to the existing mobile payment application and only needs one step (touch) to pay the transaction, followed by verification and authorisation. Hence, QRIS is effortless and far from complicated in terms of ease of use.

Another evidence of our study is social influence (SI) was significantly determined intention to use QRIS. This result is in accordance with Morosan and DeFranco (2016), who exhibits a direct effect of social influence or subjective norm on the intention to use near field communication mobile payments (NFC-MP), as well as Lee (2009) for internet banking. This finding was also confirmed in the earlier study of internet banking (Rahi et al., 2018), mobile commerce (Shaw and Sergueeva, 2019) and digital wallet (Widodo et al., 2019).

According to the result of our empirical study, facilitating condition (FC) was found to significantly influence the intention to use QRIS; which was corroborated by Venkatesh et al. (2003) and other studies, for example, Shaw and Sergueeva (2019) and Widodo et al. (2019), although the effect size is relatively small. These findings could provide policymakers and payment instrument issuers with helpful feedback to boost the product's acceptance rate.

In addition, our findings indicate that the predictors of the UTAUT (Venkatesh et al., 2003) included in our proposed model are still accurate in predicting the behavioural intention toward new technology adoption, specifically for the use of QRIS in Indonesia. Even with the newly added constructs, the significance of all UTAUT latent variables supports this result, although the effect size may vary. The unique characteristic of QRIS compared to other payment instruments typically explored in prior studies and the institution's credibility factor that introduced the new technology may account for these unconventional findings in this study employing the UTAUT model.

**Theoretical Contribution**

Our study successfully differentiated the effect of government intervention (LE), COVID-19 pandemic (LE), and common risks associated with technology adoption (PR), which is absent in the previous study of technology adoption in the context of COVID-19 pandemic.

This study reveals, with some extension, that UTAUT was sufficiently demonstrated as a reliable method for assessing people's intention to adopt new technology, even in the complicated context of pandemics and government policies, particularly for payment instruments in Indonesia. Another improvement is the alteration of the patterns of effect between variables within the model, which is shown by the additional latent variables, namely FM and LE, that outperformed the conventional determinants of UTAUT. As a result, this study contributes to the development of technology acceptance theory in many aspects.

**Practical Implication**

Our empirical research reveals that central bank policy influenced QRIS adoption more than the pandemic and other UTAUT variables; hence, we advocate this measurement approach can be used used as a complementary tool to analyse the effectiveness of the central bank's policies in changing people's behaviour. As noted by Coglianese (2012), examining the audience's behaviour following policy implementation is essential for measuring the success of a regulatory institution, as people's behavioural change substantially influences the final objective of the public institution. It also corresponds with Kowalkiewicz and Dootson's (2019) recommendation that understanding people's behaviour has become increasingly important for public organisations in the digital era.

Moreover, in response to the findings that central bank policy, pandemic, and facilitating conditions are the top three most influential variables affecting users' adoption of QRIS, it is suggested that increasing public awareness of the benefits of the new technology through public campaigns and cultivating a supportive environment, as suggested by Verplanken and Wood (2006), may improve the policy's success rate.

**Limitations and Future Research**

The are several limitations that need to be acknowledged. Firstly, the QR standard is defined specifically for Indonesia, which may differ in other countries. Therefore, future studies that want to replicate this model for a similar technology product and services in other countries should consider its specification. Secondly, though it has good size, the sample cannot represent all provinces and ensure demographic classification due to time and budget constraints. Accordingly, expanding the sample sizes, especially for merchant participants and the geographical location, is suggested for a better generalisation of the study, as well as to investigate the effect of demographic characteristics on behaviour intention to use QRIS. Finally, the data collection method was a cross-sectional study restricted to Indonesia during the COVID-19 outbreak; therefore, examining the evolution of user behaviour over time is inconceivable. Furthermore, the results may not be generalised to different countries with various conditions, primarily due to the diverse impact of the COVID-19 pandemic.

Further research with the extension of UTAUT or a more comprehensive model to accommodate other possible key determinants of new technology acceptance using other new technology products may be an exciting topic to explore.

## CONCLUSION

This study examines the effect of Bank Indonesia's policy on people's behavioural intention to adopt new technology during the COVID-19 pandemic by incorporating new variables into the UTAUT model (Venkatesh et al., 2003). We added new latent variables, namely law enforcement (LE), force majeure (FM), and perceived risks (PR), to capture the effect of central bank policy and to distinguish its effect from the covid-19 pandemic on people's behaviour in adopting new payment platform namely QRIS; and also separate the risk of the pandemic from the typical risk associated with the adoption of new technologies.

This study concludes that the central bank's policies have a considerable impact on people's intention to use a new payment platform, and the measure of the policy effect can be separated from the effect of the COVID-19 pandemic (FM). We also successfully isolate the effect of pandemic risk from common risks (PR) associated with the use of new technologies and other factors influencing the adoption of new technologies. This study found that the predictors of the UTAUT (Venkatesh et al., 2003) used in our proposed model are still reliable to predict the behavioural intention toward new technology adoption even with the addition of new variables to capture the effect of government policy and unprecedented events like a pandemic. Finally, this research contributes significantly to the literature on technology acceptance and aids policymakers in assessing the significance of their policy and optimising public acceptability of new technology products or services, especially in the setting of an unprecedented scenario such as the COVID-19 pandemic.

# ACKNOWLEDGEMENT

**Informed Consent Statement:** Informed consent was obtained from all participants involved in the study.

**Acknowledgements:** The author would like to thank all those who helped me during this research project. The author would like to express gratitude to his supervisors, Prof. John Fender, DR. Yiannis Karavias, and DR. Christoph Gortz, Head of Payment System Policy Department of Bank Indonesia, and the online survey participants.

**Funding:** This work was supported by the LPDP Indonesia.

**Conflicts of Interest:** The authors declare no conflict of interest.

*Appendix*

# Question of indicators and latent variables

| Factors | Indicators |
|---|---|
| **Performance Expectancy (PE)** | − I find QRIS (QR code Indonesian Standard) useful in my daily life. |
| | − Using QRIS (QR code Indonesian Standard) helps me accomplish things more quickly. |
| | − Using QRIS (QR code Indonesian Standard) increases my productivity. |
| **Effort Expectancy (EE)** | − Learning how to use QRIS (QR code Indonesian Standard) is easy for me. |
| | − My interaction with QRIS (QR code Indonesian Standard) is clear and understandable. |
| | − I find QRIS (QR code Indonesian Standard) easy to use. |
| | − It easy for me to become skilful at using QRIS (QR code Indonesian Standard). |
| **Social Influence (SI)** | − People who influence my behaviour think that I should use QRIS (QR code Indonesian Standard). |
| | − People who are important to me think that I should use QRIS (QR code Indonesian Standard). |
| | − People whose opinions that I value prefer that I use QRIS (QR code Indonesian Standard). |
| **Facilitating Conditions (FC)** | − I have the resources necessary to use QRIS (QR code Indonesian Standard). |
| | − I have the knowledge necessary to use QRIS (QR code Indonesian Standard). |
| | − QRIS (QR code Indonesian Standard) is compatible with other technologies I use. |
| | − I can get help from others when I have difficulties using QRIS (QR code Indonesian Standard). |
| **Perceived Risk (PR)** | − I am worried that my personal information can be stolen in the transaction using QRIS (QR code Indonesian Standard). |
| | − I think using QRIS (QR code Indonesian Standard) can cause me to lose money, or my transaction might be altered by someone else. |
| | − Overall, I believe that the overall risks of QRIS (QR code Indonesian Standard) are high. |
| **Force Majeure (FM)** | − I think using QRIS (QR code Indonesian Standard) is safer (in this pandemic period). |
| | − I think using QRIS (QR code Indonesian Standard) is necessary to avoid jeopardy. |
| **Law Enforcement (LE)** | − I was suggested by the authority to use QRIS (QR code Indonesian Standard) for the particular transaction. |
| | − I think the authority's suggestion to use QRIS (QR code Indonesian Standard) was beneficial to me. |
| **Behavioural Intention (BI)** | − I intend to continue using electronic money in the future. |
| | − I will always try to use QRIS (QR code Indonesian Standard) in my daily life. |
| | − I plan to continue to use QRIS (QR code Indonesian Standard) frequently. |

# Figures and Tables

Figure 1. Research Model

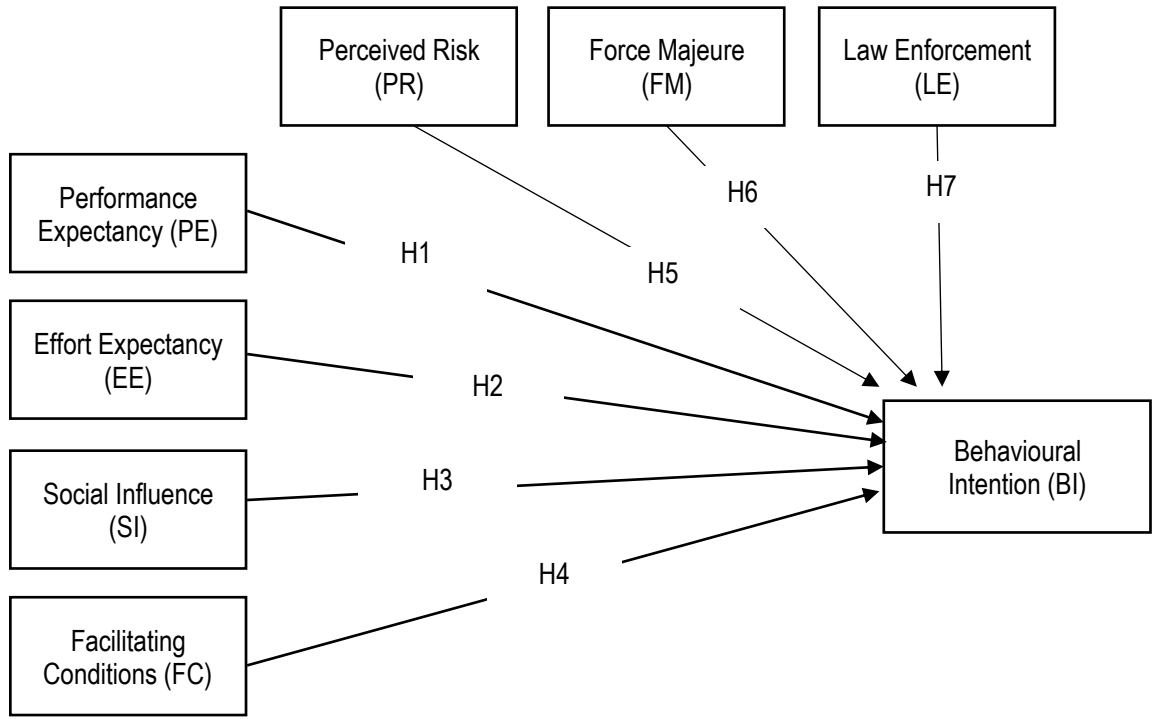

Table 1. Respondent's Demographic Profile

| Variable | Description | N | (%) |
|---|---|---|---|
| Gender | Male | 239 | 41.78% |
|  | Female | 333 | 58.22% |
| Age | 18 – 25 | 165 | 28.85% |
|  | 26 – 35 | 171 | 29.90% |
|  | 36 – 45 | 150 | 26.22% |
|  | 46 – 55 | 81 | 14.16% |
|  | >55 | 5 | 0.87% |
| Educational Level | Junior High school/Primary Edu | 4 | 0.70% |
|  | High school/equivalent | 105 | 18.36% |
|  | Diploma | 35 | 6.12% |
|  | S1 | 364 | 63.64% |
|  | S2/S3 | 64 | 11.19% |
| Marital status | single (unmarried) | 226 | 39.51% |
|  | married | 335 | 58.57% |
|  | divorced | 9 | 1.57% |
|  | widowed | 2 | 0.35% |
| Location | Jakarta capital city and surrounding (Jabodetabek) | 168 | 29.37% |
|  | Province Capital City | 207 | 36.19% |
|  | Outside province cap.city | 197 | 34.44% |
| Internet access | Available | 475 | 83.04% |
|  | Not available | 97 | 16.96% |
| Duration of Use QRIS | < 3 Months | 87 | 15.21% |
|  | 3 – 6 Months | 84 | 14.69% |
|  | 6 Months – 1 Year | 111 | 19.41% |
|  | >1 Years | 290 | 50.70% |

Source: Data processing

Table 2. The Result of Reliability and Convergence Validity

| Factors | Items | Loadings | CA | CR | AVE |
|---|---|---|---|---|---|
| Performance expectation | PE1 | 0.876 | 0.842 | 0.905 | 0.760 |
| | PE2 | 0.887 | | | |
| | PE3 | 0.853 | | | |
| Effort Expectancy | EE1 | 0.888 | 0.933 | 0.952 | 0.833 |
| | EE2 | 0.928 | | | |
| | EE3 | 0.930 | | | |
| | EE4 | 0.905 | | | |
| Social Influence | SI1 | 0.938 | 0.922 | 0.951 | 0.866 |
| | SI2 | 0.950 | | | |
| | SI3 | 0.903 | | | |
| Facilitating Condition | FC1 | 0.870 | 0.876 | 0.924 | 0.801 |
| | FC2 | 0.914 | | | |
| | FC3 | 0.901 | | | |
| Perceived Risk | PR1 | 0.884 | 0.900 | 0.957 | 0.881 |
| | PR2 | 0.924 | | | |
| | PR3 | 0.927 | | | |
| Force Majeure | FM1 | 0.922 | 0.813 | 0.915 | 0.843 |
| | FM2 | 0.914 | | | |
| Law Enforcement | LE1 | 0.887 | 0.794 | 0.905 | 0.827 |
| | LE2 | 0.932 | | | |
| Behavioural Intention | BI1 | 0.931 | 0.932 | 0.957 | 0.881 |
| | BI2 | 0.937 | | | |
| | BI3 | 0.947 | | | |

*Note: CA, Cronbach's alpha; CR, composite reliability; AVE, average variance extracted*

Table 3. Discriminant Validity, Fornell and Larcker (1981) Criterion

| | BI | EE | FC | FM | LE | PR | PE | SI |
|---|---|---|---|---|---|---|---|---|
| **BI** | 0.938 | | | | | | | |
| **EE** | 0.632 | 0.913 | | | | | | |
| **FC** | 0.614 | 0.729 | 0.895 | | | | | |
| **FM** | 0.661 | 0.593 | 0.527 | 0.918 | | | | |
| **LE** | 0.656 | 0.509 | 0.498 | 0.589 | 0.910 | | | |
| **PR** | -0.285 | -0.274 | -0.241 | -0.266 | -0.212 | 0.912 | | |
| **PE** | 0.632 | 0.634 | 0.536 | 0.613 | 0.569 | -0.222 | 0.872 | |
| **SI** | 0.628 | 0.519 | 0.461 | 0.547 | 0.581 | -0.178 | 0.623 | 0.930 |

Table 4. Validity testing, Heterotrait-Monotrait Ratio (HTMT) Criterion

|    | BI    | EE    | FC    | FM    | LE    | PR    | PE    | SI |
|----|-------|-------|-------|-------|-------|-------|-------|----|
| **BI** |       |       |       |       |       |       |       |    |
| **EE** | 0.677 |       |       |       |       |       |       |    |
| **FC** | 0.679 | 0.805 |       |       |       |       |       |    |
| **FM** | 0.758 | 0.680 | 0.624 |       |       |       |       |    |
| **LE** | 0.753 | 0.580 | 0.588 | 0.723 |       |       |       |    |
| **PR** | 0.305 | 0.293 | 0.265 | 0.305 | 0.236 |       |       |    |
| **PE** | 0.713 | 0.715 | 0.623 | 0.740 | 0.682 | 0.253 |       |    |
| **SI** | 0.678 | 0.559 | 0.511 | 0.632 | 0.674 | 0.193 | 0.706 |    |

Table 5. Structural Model Hypothesis Testing

| Hypothesis | Relationship | Path Coeff | t-value | Decision | f2 | Q2 | 95% CI LL | 95% CI UL |
|---|---|---|---|---|---|---|---|---|
| H1 | Performance Expectation -> Behavioural Intention | 0.096 | 2.401* | Supported | 0.012 | 0.008 | 0.019 | 0.175 |
| H2 | Effort Expectancy -> Behavioural Intention | 0.096 | 2.081* | Supported | 0.009 | 0.008 | 0.004 | 0.186 |
| H3 | Social Influence -> Behavioural Intention | 0.185 | 3.556** | Supported | 0.049 | 0.035 | 0.088 | 0.289 |
| H4 | Facilitating Condition -> Behavioural Intention | 0.173 | 4.217** | Supported | 0.037 | 0.026 | 0.092 | 0.254 |
| H5 | Perceived Risk -> Behavioural Intention | -0.060 | 2.614** | Supported | 0.009 | 0.005 | -0.108 | -0.016 |
| H6 | Force Majeure -> Behavioural Intention | **0.203** | 4.363** | Supported | 0.057 | 0.040 | 0.113 | 0.295 |
| H7 | Law Enforcement -> Behavioural Intention | **0.227** | 4.465** | Supported | 0.074 | 0.053 | 0.126 | 0.325 |

Note(s): **Significant at p-value < 0.01, *significant at p-value < 0.05